\documentclass[twocolumn]{aastex62}

\newcommand{\um}{$\mu$m}

\accepted{}

\shorttitle{The dusty torus of NGC 1068}
\shortauthors{Lopez-Rodriguez et al.}

\begin{document}

\title{The emission and distribution of dust of the torus of NGC 1068}

\correspondingauthor{Lopez-Rodriguez, E.}
\email{elopezrodriguez@sofia.usra.edu}

\author{Enrique Lopez-Rodriguez}
\affil{SOFIA Science Center, NASA Ames Research Center, Moffett Field, CA 94035, USA}
\affil{National Astronomical Observatory of Japan, 2-21-1 Osawa, Mitaka, Tokyo 181-8588, Japan}


\author{Lindsay Fuller}
\affiliation{Department of Physics and Astronomy, University of Texas at San Antonio, One UTSA Circle, San Antonio, TX 78249, USA}

\author{Almudena Alonso-Herrero}
\affiliation{Centro de Astrobiolog\'ia (CAB, CSIC-INTA), ESAC Campus, E-28692 Villanueva de la Ca\~nada, Madrid, Spain}
\affiliation{Department of Physics and Astronomy, University of Texas at San Antonio, One UTSA Circle, San Antonio, TX 78249, USA}

\author{Andreas Efstathiou}
\affiliation{School of Sciences, European University Cyprus, Diogenes Street, Engomi, 1516 Nicosia, Cyprus}

\author{Kohei Ichikawa}
\affiliation{Department of Astronomy, Columbia University, 550 West 120th Street, New York, NY 10027, USA}
\affiliation{National Astronomical Observatory of Japan, 2-21-1 Osawa, Mitaka, Tokyo 181-8588, Japan}
\affiliation{Department of Physics and Astronomy, University of Texas at San Antonio, One UTSA Circle, San Antonio, TX 78249, USA}

\author{Nancy A. Levenson}
\affiliation{Space Telescope Science Institute, 3700 San Martin Dr, Baltimore, MD 21218, USA}

\author{Chris Packham}
\affiliation{Department of Physics and Astronomy, University of Texas at San Antonio, One UTSA Circle, San Antonio, TX 78249, USA}

\author{James Radomski}
\affil{SOFIA Science Center, NASA Ames Research Center, Moffett Field, CA 94035, USA}

\author{Cristina Ramos Almeida}
\affiliation{Instituto de Astrof\'isica de Canarias, C/V\'ia L\'actea, s/n, E-38205 La Laguna, Tenerife, Spain}
\affiliation{Departamento de Astrof\'isica, Universidad de La Laguna, E-38205 La Laguna, Tenerife, Spain}


\author{Dominic J. Benford}
\affil{Goddard Space Flight Center, Greenbelt, MD 20771 USA}

\author{Marc Berthoud}
\affil{Yerkes Observatory, Williams Bay, WI}

\author{Ryan Hamilton}
\affil{SOFIA Science Center, NASA Ames Research Center, Moffett Field, CA 94035, USA}

\author{Doyal Harper}
\affil{Yerkes Observatory, Williams Bay, WI}

\author{Attila Kov\'avcs}
\affil{California Institute of Technology, 301-17, 1200 East California Blvd, Pasadena, CA 91125, USA}

\author{Fabio P. Santos}
\affil{Center for Interdisciplinary Exploration and Research in Astrophysics (CIERA) and Department of Physics \& Astronomy, Northwestern University, 2145 Sheridan Road, Evanston, IL 60208, U.S.A.}

\author{J. Staguhn}
\affil{NASA Goddard Space Flight Center, Code 665, Greenbelt, MD 20771, USA}
\affil{Department of Physics \& Astronomy, Johns Hopkins University, Baltimore, MD, 21218, USA}


\author{Terry Herter}
\affil{Astronomy Department, 202 Space Sciences Building, Cornell University, Ithaca, NY 14853-6801, USA}



\begin{abstract}

We present observations of NGC 1068 covering the $19.7-53.0$ \um~wavelength range using FORCAST and HAWC+ onboard SOFIA. Using these observations, high-angular resolution infrared (IR) and sub-mm observations, we find an observational turn-over of the torus emission in the $30-40$ \um~wavelength range with a characteristic temperature of $70-100$ K. This component is clearly different from the diffuse extended emission in the narrow line and star formation regions at 10-100 \um~within the central 700 pc. We compute $2.2-432$ \um~2D images using the best inferred \textsc{clumpy} torus model based on several nuclear spectral energy distribution (SED) coverages. We find that when $1-20$ \um~SED is used, the inferred result gives a small torus size ($<4$ pc radius) and a steep radial dust distribution. The computed torus using the $1-432$ \um~SED provides comparable torus sizes, $5.1^{+0.4}_{-0.4}$ pc radius,  and morphology to the recently resolved  432 \um~ALMA observations. This result indicates that the $1-20$ \um~wavelength range is not able to probe the full extent of the torus. The characterization of the turn-over emission of the torus using the $30-60$ \um~wavelength range is sensitive to the detection of cold dust in the torus. The morphology of the dust emission in our 2D image at 432 \um~is spatially coincident with the cloud distribution, while the morphology of the emission in the $1-20$ \um~wavelength range shows an elongated morphology perpendicular to the cloud distribution. We find that our 2D \textsc{clumpy} torus image at 12 \um~can produce comparable results to those observed using IR interferometry.
\end{abstract}




\section{Introduction} \label{sec:intro}

NGC 1068 \citep[D = 14.4 Mpc, ][and 1\arcsec~= 70 pc, adopting H$_{0}$ = 73 km s$^{-1}$ Mpc$^{-1}$]{Bland-Hawthorn:1997aa} is the archetypical type 2 active galactic nucleus (AGN). The emission from its central engine is obscured by a distribution of optically thick dust. This dusty distribution has recently \citep{Gallimore:2016ab,Garcia-Burillo:2016aa,2018arXiv180106564I} been resolved with the Atacama Large Millimeter Array (ALMA) observations using continuum and emission line observations with angular resolution $<0.1\arcsec$~($<7$ pc), which have provided tight constraints on the torus size and morphology. Specifically, \citet{Garcia-Burillo:2016aa} measured a $7-10$ pc torus diameter using 432 \um~continuum emission, \citet{Gallimore:2016ab} found a $12 \times 7$ pc structure through the study of the CO J = 6 $\rightarrow$ 5 emission, and \citet{2018arXiv180106564I} found that both the morphology and dynamics of the HCN J=3$\rightarrow$2 and HCO$^{+}$ J=3$\rightarrow$2 emission are fairly aligned in the east-west direction with a size of $\sim$12$\times$5 pc. In addition, infrared (IR) interferometric observations \citep{Wittkowski:2004aa,Jaffe:2004aa,Raban:2009aa,Lopez-Gonzaga:2014aa} have put tight constraints in the different emission components of the torus in NGC 1068. Specifically, a 700K dust structure with a size 1.4 pc, 250K structure with a size of 3 pc, and an extended 14 pc dust component in the polar direction with a characteristic temperature of 350K. These observations have challenged our current understanding on the emission and distribution of dust surrounding the active nucleus of NGC 1068.

The torus is not resolved by the current suite of single-dish telescopes, thus SED modeling using the best angular resolution to isolate the torus emission from extended diffuse dust emission, star formation regions and/or host galaxy is crucial to obtain physical information about the torus \citep[i.e.][]{Mason:2006aa,Honig:2008aa,Ramos-Almeida:2011aa,Alonso-Herrero:2011aa,Feltre:2012aa,Ramos-Almeida:2014aa, Ichikawa:2015aa}.  These extensive works have been performed in the 1$-$20 \um~wavelength range with the general agreement that the torus is formed by a clumpy distribution of optically thick dust with sizes of few pc surrounding the central engine. The 1$-$20 \um~high-angular resolution observations show an increase in the total flux density with increasing wavelength. The silicate feature at 10 \um~and 18 \um, the near-IR (NIR) emission, and the luminosity of the torus in the 10 \um~window provide important diagnostic tools to constrain the torus structure \citep{2017NatAs...1..679R}. However, these studies show that the turn-over of the torus emission occurs in the 20$-$30 \um, which makes the torus emission to be dominated by warm dust with a characteristic temperature of $\sim$100-150K, and with typical torus diameters of $\le$5 pc--slightly smaller than the currently resolved observations by ALMA.

There is an observational gap within the $20-70$ \um~wavelength range with angular resolutions  $<10$\arcsec, where warm/cold dust in the torus seems to have its peak of emission, and that it has not been characterized. A Bayesian exploration study by \citet{Asensio-Ramos:2013aa} using \textsc{clumpy} torus models found that the region between 10 and 200 \um~provides the best wavelength range to constrain the torus radial extent and the number and radial distribution of clouds in the torus. \citet{Fuller:2016aa} observationally show the potential of far-IR (FIR) observations to constrain the torus size using 31.5 \um~imaging observations with the Faint Object Infrared Camera for the SOFIA Telescope (FORCAST) onboard of SOFIA. They found that 1) the torus radial extent model parameter decreases by a 30\% in size for 60\% (6 out of 10 AGN) of their sample, and 2) the SED turn-over of the torus emission does not occur up to 31.5 \um, in $F_{\nu}$. Their observations also show resolved diffuse extended emission along the narrow line region (NLR), which allowed them to better isolate the torus emission using SOFIA than previous \textit{Spitzer} $30-40$ \um~observations. The combination of fully sampled nuclear SED, resolved IR interferometric and ALMA observations, and torus models is crucial to break degeneracies in the physical properties of the torus.

With the tight constrain in the torus size of NGC 1068 provided by the resolved images by ALMA, and the currently available moderate angular resolution FIR capabilities, we here present an observational study to characterize the emission and distribution of dust through the characterization of the SED of NGC 1068 using torus models. We present observations of NGC 1068 covering the $19.7-37.1$ \um~wavelength range using FORCAST and newly obtained 53.0 \um~imaging observations by the High-resolution Airborne Wideband Camera (HAWC+) onboard SOFIA. The paper is organized as follows: Section \ref{sec:obs} describes the observations and data reduction, Section \ref{sec:tor} discusses the emission and dust distribution of dust in the torus of NGC 1068, and Section \ref{sec:seddes} shows the spectral decomposition of the nuclear SED. In Section \ref{sec:con} we present our conclusions.


\section{Observations and data reduction} \label{sec:obs}

\subsection{FORCAST observations \label{subsec:for}}

NGC 1068 was observed as part of the Guaranteed Time Observations (GTO; PI: Herter, T.) on 2016 September 17 using FORCAST \citep{Herter:2012aa} on the 2.5-m SOFIA telescope. We made observations with the dual-channel mode at the 19.7 \um, 31.5 \um~and 37.1 \um~using the two-position chop-nod (C2N) method with symmetric nod-match-chop (NMC) to remove time-variable sky background and telescope thermal emission and to reduce the effect of 1/\textit{f} noise from the array. In all observations, we used an instrumental position angle, i.e. long-axis of the detector with respect to the North on the sky, of 305$^{\circ}$, a chop-throw of 1\arcmin~with a 30$^{\circ}$ E of N chop-angle. The on-source times were 427s, 471s, and 343s at 19.7 \um, 31.5 \um~and 37.1 \um, respectively.

SOFIA provided reduced data using the \textsc{forcast redux pipeline v1.1.3} following the method described by \citet{Herter:2013aa} to correct for bad pixels, ``droop" effect, non-linearity, and cross-talk. The point spread functions (PSFs) of the observations were estimated using observations of Ceres taken immediately before NGC 1068 observations with the same instrumental configuration and bands. We estimated a full-width at the half maximum (FWHM) of Ceres of 2.4\arcsec, 2.8\arcsec~and 2.9\arcsec~at 19.7 \um, 31.5 \um~and 37.1 \um, respectively. NGC 1068 was flux-calibrated using the set of standard stars of the observing run, which provides flux uncertainties of 5.0\%, 5.2\% and 7.7\% at 19.7 \um, 31.5 \um~and 37.1 \um, respectively.

\subsection{HAWC+ observations \label{subsec:haw}}

\begin{figure}[ht!]
\includegraphics[angle=0,scale=0.43]{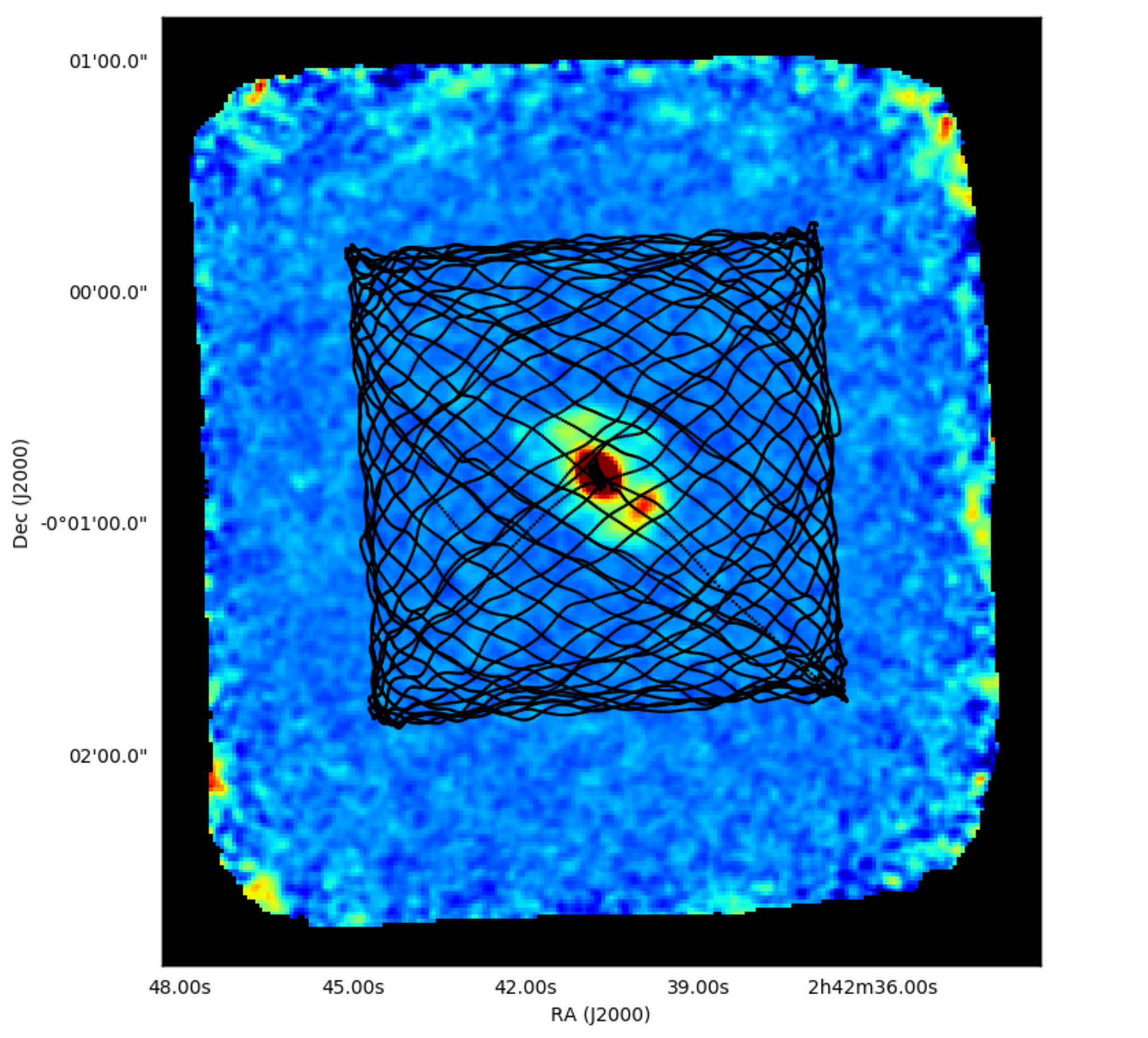}
\caption{Example of a single Lissajous pattern (black line) overlaid on the final image at 53 \um~of NGC 1068 (color scale) using several Lissajous scans.
\label{fig:fig1}}
\epsscale{2.}
\end{figure}

NGC 1068 was observed as part of the GTO (PI: Dowell, D.) on 2017 May 06 using HAWC+ \citep[][Harper et al. in preparation]{Vaillancourt:2007aa} on the 2.5-m SOFIA telescope. We made observations using the Lissajous pattern in the total intensity mode at 53 \um~($\lambda_{c} =$ 53 \um, $\Delta\lambda/\lambda_{c} =$ 0.17 bandwidth). In this new SOFIA observing mode, the telescope is driven to follow a parametric curve at a non-repeating period whose shape is characterized by the relative phases and frequency of the motion. Fig. \ref{fig:fig1} shows the Lissajous pattern of a single observation at 53 \um~of NGC 1068 with a scan rate of 100\arcsec~s$^{-1}$ and a 60\arcsec~scan amplitude. We performed a total of five Lissajous scans with relative phases of 5$^{\circ}$ and 27$^{\circ}$ with a total on-source time of 455s.


We reduced the data using the Comprehensive Reduction Utility for SHARC II v2.34-3beta \citep[\textsc{crush}, ][]{Kovacs:2006aa,Kovacs:2008aa}\footnote{\textsc{crush} can be found at: \url{http://www.submm.caltech.edu/~sharc/crush/}} optimized for HAWC+ and the \textsc{hawc\_dpr pipeline v1.1.1}. \textsc{crush} estimates and removes the correlated atmospheric and instrumental signals, solves for the relative detector gains, and determines the noise weighting of the time streams in an iterated pipeline scheme. The PSF was estimated using Uranus observations on 2017 May 07 with a FWHM of 4.9\arcsec, consistent with diffraction-limited observations at 53 \um. Flux calibrators were not observed during the same flight, thus we cross-calibrated our observations using flux calibrators, i.e. Uranus, from other flights. Although we find a flux calibration accuracy of $\sim$8\% for observations taken within the same flight, the cross-calibration between flights can only ensure a flux accuracy of $\sim$20\%.

\subsection{Photometry and nuclear flux imaging modeling  \label{subsec:pho}}

\begin{figure*}[ht!]
\includegraphics[angle=0,scale=0.34]{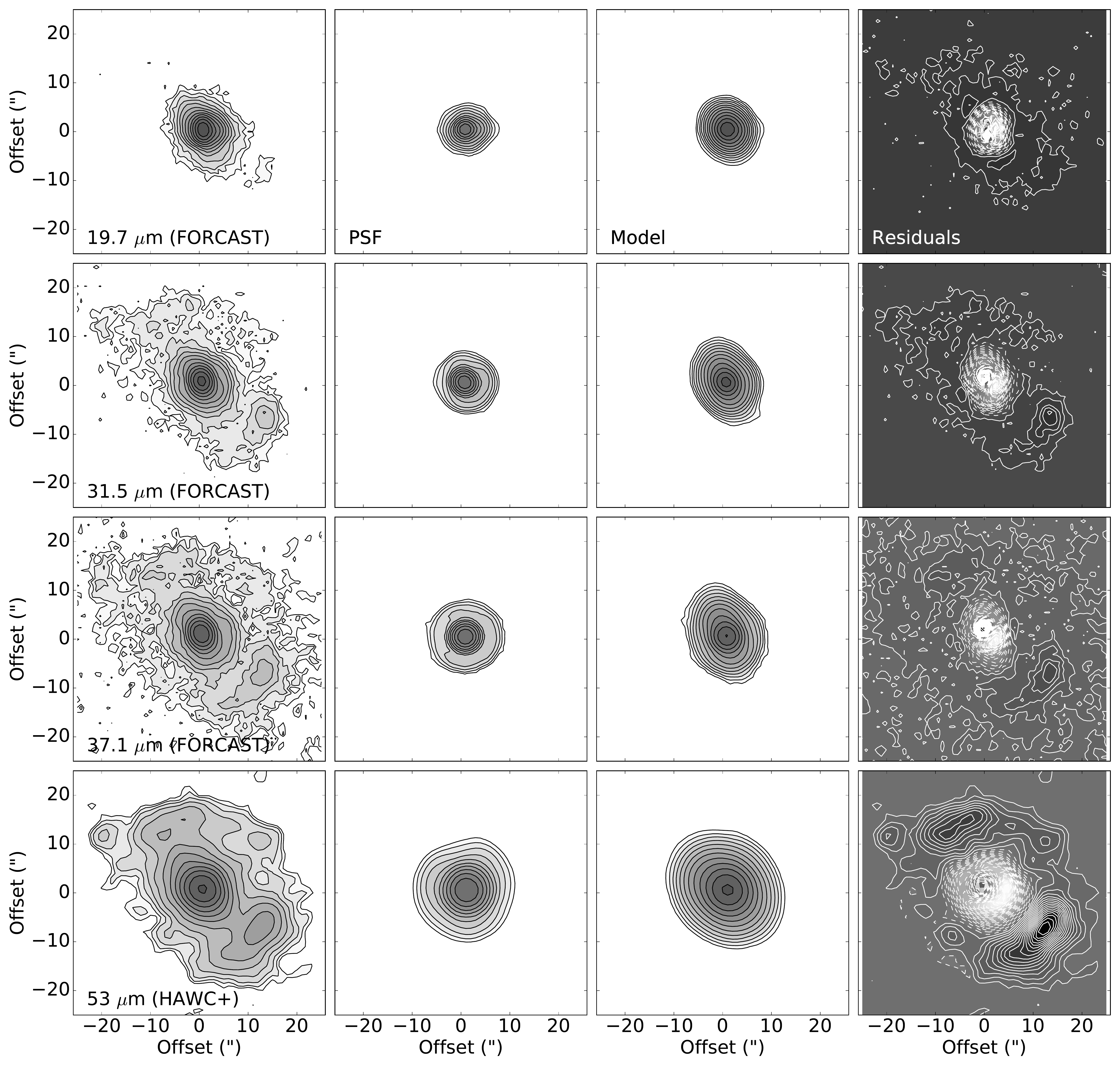}
\caption{From left to right: NGC 1068 observations, scaled PSF, Model (PSF+2D Gaussian) of the central $20\arcsec  \times 20\arcsec$  (1.4 $\times$ 1.4 kpc$^{2}$), and Residuals (NGC1068$-$Model) at 19.7 \um, 31.5 \um, 37.1 \um~and 53 \um~from top to bottom, respectively. In all cases, the FOV is $50\arcsec \times 50\arcsec$ (3.5 $\times$ 3.5 kpc$^{2}$). NGC1068, PSF and Model contours are shown in $\log{\mbox{(flux density [Jy])}}$ from -2.0 to 1.5 in steps of 0.2. Residual contours are shown in flux density (Jy) from -0.4  Jy to 0.3 Jy in steps of 0.02 Jy. North is up and East is left.
\label{fig:fig2}}
\epsscale{2.}
\end{figure*}

We aim to obtain the emission from the unresolved core of NGC 1068 at all the observed wavelengths. The SOFIA observations of NGC 1068 shows a resolved core (Figure \ref{fig:fig2}) which is thought to arise from an unresolved and an extended components. To obtain the fractional contribution to the total emission from both unresolved and extended components, we made two different photometric measurements (Table \ref{tab:table1}). First, the flux in a circular aperture of 10\arcsec~(700 pc) diameter was measured, which ensures to enclose the whole flux of an unresolved source at the given wavelength and minimizes the contribution from the diffuse extended emission. Second, the central 20\arcsec~$\times$ 20\arcsec~(1.4 $\times$ 1.4 kpc$^{2}$) emission was fitted with a composite model using the corresponding PSF to each observation and a 2D Gaussian profile. We refer to these methods as ``aperture'' and ``PSF-scaling'' photometry, respectively, in the remainder of the paper. The aperture photometry represents the total flux from the observed galaxy at a given wavelength, F$_{T}$. In the PSF-scaling method, the total flux from the scaled-PSF, F$_{PSF}$, represents the maximum likely contribution from an unresolved nuclear component at the given angular resolution of the observations, while the total flux from the 2D Gaussian profile, F$_{ext}$, provides the minimum contribution of the extended component surrounding the central source. We estimated the total flux of the model, F$_{T}^{M}$, as the sum of both the PSF and the 2D Gaussian profile. The PSF-scaling method has five free parameters, the amplitudes of both PSF and 2D Gaussian profile, the FWHM of the long, $b$, and short, $a$, axis and the position angle (P.A.) of the 2D Gaussian profile. The fitting routine minimizes the residuals (galaxy minus model: scaled-PSF $+$ 2D Gaussian) to a level $<$5\% of the total flux, F$_{T}$, within the central 20\arcsec~diameter. We also considered a 2D S\'ersic profile to fit the extended component. We obtained index profiles $\sim$0.5 and size parameters of the S\'ersic profiles similar to the FWHM of the observations, which is close to the special case of the S\'ersic profile tending to a Gaussian profile. Due to this behaviour and that the S\'ersic profile increases the number of free parameters, we decided to use 2D Gaussian profiles.

The uncertainty in the photometry was estimated in the following manner. The aperture photometry uses the flux uncertainties estimated from the flux calibration described in Sections \ref{subsec:for} and \ref{subsec:haw}. For the PSF-scaling photometry, an estimate of the error induced by a variable PSF was obtained by cross-calibrating the standard stars observed on the same or several nights. This error was found to be $\sim$5\%. Another estimate of the error induced by the fitting procedure was estimated to be $\sim$3\%. The total uncertainty for the PSF-scaling photometry was calculated by adding in quadrature these individual contributions. 


\begin{deluxetable*}{ccccccc|ccc}[ht]
\tablecaption{Measured and modeled nuclear photometry. Fractional contribution of emissive components in a 10\arcsec~aperture.\label{tab:table1}}
\tablecolumns{7}
\tablenum{1}
\tablewidth{0pt}
\tablehead{
\multicolumn{7}{c}{Photometric measurements\tablenotemark{a}} & \multicolumn{3}{c}{Spectral Decomposition\tablenotemark{b}}\\
\cline{8-10}
\colhead{$\lambda_{c}$} &  \colhead{F$_{T}$} & \colhead{F$_{T}^{M}$} & \colhead{F$_{PSF}$} & \colhead{F$_{ext}$} & \colhead{PSF} & \colhead{Extended}	& \colhead{Star Formation}	&	\colhead{Dust at 200K}		&	\colhead{Torus} \\
\colhead{(\um)} &  \colhead{(Jy)} & \colhead{(Jy)} & \colhead{(Jy)} & \colhead{(Jy)}  &  \colhead{\%} &  \colhead{\%} &  \colhead{\%} &  \colhead{\%} &  \colhead{\%}
}
\startdata
19.7	&	$61.9\pm3.1$	&	$60.9\pm3.8$	&	$22.0\pm1.4$	&	$38.9\pm2.4$	&	$36\pm4$		&	$64\pm7$		& 
10	&	60	&	30		\\
31.5	&	$59.4\pm3.1$	&	$58.4\pm3.7$	&	$28.8\pm1.8$	&	$29.6\pm1.9$	&	$49\pm5$		&	$51\pm6$		& 
23	&	36	&	41	\\
37.1	&	$59.6\pm4.6$	&	$58.8\pm5.0$	&	$29.7\pm2.5$	&	$29.1\pm2.5$	&	$51\pm8$		&	$49\pm8$		& 
33	&	26	&	41	\\
53.0	&	$71.6\pm14.3$	&	$71.5\pm14.5$	&	$23.8\pm4.8$	&	$47.7\pm9.7$	&	$33\pm15$	&	$67\pm25$	& 
64	&	16	&	20	\\
\enddata
\tablenotetext{a}{Measured and modeled photometry as described in Section \ref{subsec:pho}.}
\tablenotetext{b}{Fractional contribution of the several components used in the spectral decomposition described in Section \ref{sec:seddes}. We estimate a 5\% uncertainty for the fractional contribution of each component.}
\end{deluxetable*}


Figure \ref{fig:fig2} shows the NGC 1068 observations, scaled PSF, Model (PSF$+$2D Gaussian) and residuals (NGC 1068$-$Model) of the central 50\arcsec~$\times$ 50\arcsec~(3.5 $\times$ 3.5 kpc$^{2}$) observations at 19.7 \um, 31.5 \um, 37.1 \um~ and 53 \um. Table \ref{tab:table1} shows the measured and modeled nuclear photometry for each photometric method and model component in the central 10\arcsec~aperture. The fractional contribution of the PSF and 2D Gaussian profile in the central 10\arcsec~diameter is also shown. Our flux density at 19.7 \um~estimated by using PSF-scaling of $22.0\pm1.4$ Jy is in excellent agreement with the flux density of $20.2\pm3.4$ Jy in a 0.4\arcsec~aperture by \citet{Tomono:2001aa}. We took PACS/\textit{Herschel} spectroscopic data of NGC 1068 from the \textit{Herschel} Archive and we obtained a nuclear flux density of $\sim70$ Jy at 60 \um~per spaxel, where a spaxel is $9.4\arcsec \times 9.4\arcsec$ (658 $\times$ 658 pc$^{2}$). Despite the difference in wavelength, this result is in good agreement with our total flux of $72\pm14$ Jy at 53 \um~using HAWC+. At all wavelengths, the total flux model, F$_{T}^{M}$, is $<2$\% of the total flux of the observations, F$_{T}$, and the P.A. of the extended emission is $44.7^{\circ}\pm1.3^{\circ}$ with a decrease of the ratio of the short and long axis, $a/b$, from 0.85 at 19.7 \um~to 0.61 at 53 \um. Our extended diffuse emission is spatially coincident with the large scale, 32\arcsec~(1.92 kpc) inner bar at a P.A. of $48\pm3^{\circ}$, the so-called NIR bar \citep{Scoville:1988aa,Schinnerer:2000aa,Emsellem:2006aa}. Despite any contribution of diffuse extended emission within the PSF of SOFIA in the $20-53$ \um~wavelength range (Section \ref{sec:seddes}), we found a turn-over of the unresolved emission, F$_{PSF}$ in Table  \ref{tab:table1}, within the $31.5-53$ \um~wavelength range. Specifically, the PSF fractional contribution to the total flux decreases from $\sim$50\% in the $30-40$ \um~wavelength range to $<40$\% at shorter and longer wavelengths. This result is in agreement with the observational constraint by \cite{Fuller:2016aa}, who suggested that the turn-over of the torus emission does not occur until wavelengths $>31.5$ \um~for a sample of nearby AGN.


\section{The torus of NGC 1068}\label{sec:tor}

\subsection{\textsc{Clumpy} torus models}\label{sec:Ctor}


\begin{figure}[ht!]
\includegraphics[angle=0,scale=0.47]{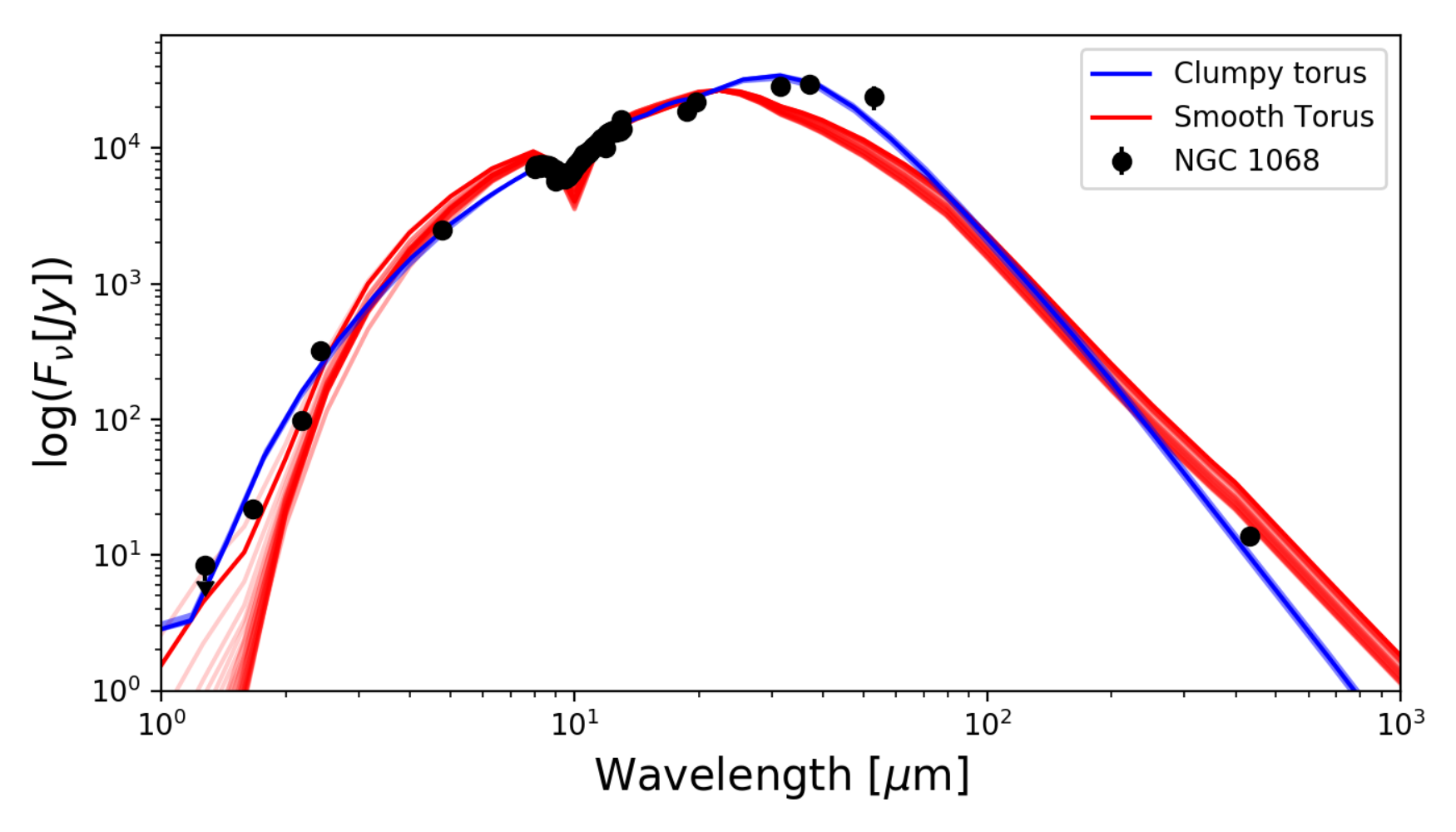}
\caption{Best fit (solid line) and 1-$\sigma$ uncertainties (shadowed area) of the \textsc{clumpy}  (blue) and smooth (red) torus models to the nuclear SED (black dots) of NGC 1068. 
\label{fig:fig3}}
\epsscale{2.}
\end{figure}


We here describe the details of the fitting to the nuclear SED of NGC 1068. The nuclear SED is composed of our PSF-scaling photometry (F$_{PSF}$) in conjunction with the 0.4\arcsec~aperture photometry from 2 \um~to 20 \um~photometry by \citet{Tomono:2001aa}, the $8-13$ \um~nuclear spectrum in a 0.4\arcsec~aperture using Michelle on the 8.1-m Gemini-North Telescope by \citet{Mason:2006aa}, and the 432 \um~ALMA observation by \citet{Garcia-Burillo:2016aa}. We fitted the nuclear SED using the \textsc{clumpy} torus models of \citet{Nenkova:2008ab} and the \textsc{BayesClumpy} approach developed by \cite{Asensio-Ramos:2009aa}. This approach has been successfully applied to this and other Seyfert galaxies \citep[e.g.][]{Alonso-Herrero:2011aa,Ichikawa:2015aa}. The free parameters of the model were set with a flat prior distribution, with the exception of the foreground visual extinction to the core, A$_{V}$, was set to be in the $0-10$ magnitude range. The best inferred model with the 1-$\sigma$ uncertainty region is shown in Figure \ref{fig:fig3}, and the posterior distributions of each model parameter are shown in Section \ref{app:pos}. Table \ref{tab:table2} shows the output values of each torus model parameter.

In general, the full family of \textsc{clumpy} torus model solutions when using the SOFIA observations from $20-53$ \um~provides a tighter 1-$\sigma$ dispersion than previously studies (i.e.  \citet{Alonso-Herrero:2011aa}, fig. 5; \cite{Garcia-Burillo:2016aa}, fig. 4, our Fig. \ref{fig:fig5}). This result is due to the better sampling of the turn-over of the torus emission in the $30-40$ \um~as in comparison with previous studies. The median value for the foreground visual extinction was found to be  A$_{V} = 9^{+1}_{-1}$ mag. We notice that if the extinction to the core was set to be negligible, A$_{V} < 5$ mag, then the fitting tends to obtain viewing angles, $i$, of the torus in the $30^{\circ}-40^{\circ}$ range, and does not fit the SED in the $1-5$ \um~wavelength range. \cite{Packham:1997aa} found that a visual extinction of 36 mag to the core of NGC 1068 can explain the absorptive polarization at 2.0 \um, compatible with the expected null polarization observed at 10 \um~by the emissive polarization of the torus found by \citet{Lopez-Rodriguez:2016aa}. Both works also found a visual extinction by the central dust lane to be $\sim8$ mag. Thus, we expect that the torus emission is extinguished by a column of dust into our LOS with a visual extinction in the range of $8-36$ mag, our computed visual extinction of A$_{V} = 9^{+1}_{-1}$ mag is in agreement within that range. We obtained a viewing angle, $i = 75^{+8}_{-4}$$^{\circ}$ compatible with the H$_{2}$0 maser observations in the central parsec of NGC 1068, which suggests a torus with an almost edge-on view, $\sim$90$^{\circ}$.

Based on the best inferred \textsc{clumpy} torus model, we can estimate torus morphological parameters as the outer radius, $r_{out} = r_{in} Y$ pc, where $r_{in}$ is the inner radius of the torus defined by the distance of the sublimation temperature of dust grains, $T$, as a function of the bolometric luminosity,  $L_{bol} $, as $r_{in} = 0.4 (L_{bol}/10^{45} \mbox{erg s}^{-1})^{0.5} (T/1500~\mbox{K})$ pc \citep{Barvainis:1987aa}, and the torus scale height as $H = r_{out} \sin \sigma$ pc. The estimated bolometric luminosity from our \textsc{clumpy} torus model, $L_{bol} = 5.02_{-0.15}^{+0.19} \times 10^{44}$ erg s$^{-1}$, yields an inner torus radius of $r_{in} = 0.28^{+0.01}_{-0.01}$ pc for dust grains at a temperature of 1500 K. Using $Y = 18^{+1}_{-1}$ and $\sigma = 43^{+12}_{-15}$$^{\circ}$, the torus radius and scale height are estimated to be $r_{out} = 5.1^{+0.4}_{-0.4}$ pc, $H = 3.5^{+1.0}_{-1.3}$ pc, respectively. Table \ref{tab:table2} summarizes these results. These results are in agreement with the resolved $7-10$ pc torus extension \citep{Garcia-Burillo:2016aa}, $12 \times 7$ pc \citep{Gallimore:2016ab}, and $12 \times 5$ pc \citep{2018arXiv180106564I} torus diameter using ALMA observations. We can estimate the visual extinction into our LOS, $A_{\rm v}^{\rm LOS}$, as $A_{\rm v}^{\rm LOS} = 1.086 N_{\rm 0} \tau_{\rm v} \exp{(-(i-90)^{2}/\sigma^{2})}$ mag. From our best \textsc{clumpy} torus model, we estimate $A_{\rm v}^{\rm LOS} = 248^{+201}_{-142}$ mag. Using the standard Galactic ratio $A_{\rm v}/N_{\rm H} = 5.23 \times 10^{-22}$ mag cm$^{2}$ \citep{Bohlin:1978aa}, we estimate a column density of $N_{\rm H} = 4.7^{+3.9}_{-2.7} \times 10^{23}$ cm$^{-2}$.


\begin{deluxetable*}{lcc|lcc}[ht]
\tablecaption{\textsc{Clumpy} and Smooth torus model parameters.\label{tab:table2}}
\tablecolumns{6}
\tablenum{2}
\tablewidth{0pt}
\tablehead{
\multicolumn{3}{c}{\textsc{clumpy} torus} & \multicolumn{3}{c}{Smooth torus} \\
\cline{1-6}
\colhead{Parameter}		&	\colhead{Symbol}	&	\colhead{Value}		&
\colhead{Parameter}		&	\colhead{Symbol}	&	\colhead{Value}	
}
\startdata
Angular width		&	$\sigma$	&	$43^{+12}_{-15}$$^{\circ}$ 	&	
Opening angle 		&	$\theta_{OA}$	&	$37^{+23}_{-8}$$^{\circ}$ \\
Radial thickness	&	$Y$		&	$18^{+1}_{-1}$	&	
Radial thickness	&	$Y_{s}$	&	$20^{+4}_{-4}$ \\
Number clouds along the equatorial plane	&	$N_{0}$	&	$4^{+2}_{-1}$ & 
-	&	-	&	-	\\
Index of the radial density profile	&	$q$	&	$0.08^{+0.19}_{-0.06}$	&
Index of the radial density profile	&	$q_{s}$	&	1 (fixed) \\
Optical depth of each cloud		&	$\tau_{v}$		&	$70^{+6}_{-14}$	&
Optical depth of the torus, LOS			&	$\tau_{v,s}	$	&	$250^{+20}_{-10}$ \\
Viewing angle		&	$i$		&	$75^{+8}_{-4}$$^{\circ}$ &
Viewing angle		&	$i_{s}$	&	$79^{+7}_{-10}$$^{\circ}$ \\
\cline{1-6} 
Inner radius		&	$r_{in}$ 		&	$0.28^{+0.01}_{-0.01}$ pc		&
				&	$r_{in,s}$		&	$0.41^{+0.05}_{-0.02}$ pc	\\
Outer radius		&	$r_{out}$		&	$5.1^{+0.4}_{-0.4}$ pc 		&
				&	$r_{out,s}$	&	$8.5^{+7.9}_{-0.7}$ pc	\\
Height			&	$H$			&	$3.5^{+1.0}_{-1.3}$ pc		&
				&	$H_{s}$		&	$4.2^{+0.5}_{-0.2}$ pc		\\
Bolometric luminosity (erg s$^{-1}$)&	$L_{bol}$		&	$5.02^{+0.15}_{-0.19} \times 10^{44}$&
				&	$L_{bol,s}$	&	$1.11^{+0.28}_{-1.23} \times 10^{44}$ \\
\enddata
\end{deluxetable*}


\subsection{Smooth torus models}\label{sec:Stor}

We have also used smooth torus models \citep{Efstathiou:1995ab} to fit the nuclear SED of NGC 1068. The best fit model and the output parameters are shown in Figure \ref{fig:fig3} and Table \ref{tab:table2}, respectively. In general, the smooth torus models reproduce well the nuclear SED of NGC 1068, except for the FIR ($20-60$ \um) wavelength range. In this spectral range, the smooth torus models underestimate the measured nuclear fluxes. We note that if we force these models to fit the FIR range, then the smooth torus models over-predict the sub-mm fluxes by a factor of 10 or more, and the torus size increases to a few tens of pc. Thus, we use the MIR spectroscopic observations, i.e. the 10-\um~silicate feature, and the ALMA observations to find the best fit of the smooth torus model. 

We find a smooth torus with similar physical characteristics as the \textsc{clumpy} torus (Table \ref{tab:table2}), except for the outer radius, which is larger in the case of the smooth torus. This difference is mainly due to the sublimation temperature used by both models, the smooth torus models use a maximum temperature of dust grains of 1000 K, in comparison with the 1500 K used by the \textsc{clumpy} torus models. Although we exclusively used the smooth torus models of \citet{Efstathiou:1995ab}, we speculate that other smooth torus models \cite[i.e.][]{2005A&A...437..861S,2006MNRAS.366..767F}  may similarly fail to account for the complete SED of the torus. This may be due to the fact that smooth models generally have much less flexibility in the specification of the distribution of the dust in the torus and especially its outer part to which the far-infrared and sub-mm observations are more sensitive. This certainly merits further study and we plan to pursue this in future work.


\begin{deluxetable*}{cccccccllcc}[ht]
\tablecaption{\textsc{clumpy} torus model parameters as a function of SED coverage.\label{tab:table3}}
\tablecolumns{8}
\tablenum{3}
\tablewidth{0pt}
\tablehead{
\colhead{$\sigma$ ($^{\circ}$)} &  \colhead{Y} & \colhead{N$_{0}$} & \colhead{q} & \colhead{$\tau_{v}$} & \colhead{i ($^{\circ}$)} & \colhead{r$_{out}$ (pc)} & \colhead{SED} 
}
\startdata
20$^{+5}_{-3}$   &   13$^{+4}_{-3}$  &   11$^{+2}_{-3}$   &   0.22$^{+0.20}_{-0.13}$   &   28$^{+10}_{-6}$   &   75$^{+4}_{-6}$   &  3.5$^{+1.3}_{-0.9}$  &   1$-$20 \um~SED+MIR Spectroscopy \\
31$^{+20}_{-8}$   &   19$^{+1}_{-1}$   &   5$^{+3}_{-2}$   &   0.06$^{+0.08}_{-0.04}$   &   59$^{+16}_{-13}$   &  71$^{+5}_{-3}$   & 5.5$^{+0.4}_{-0.4}$  &   1$-$20 \um~SED+MIR Spectroscopy+ALMA \\
\enddata
\end{deluxetable*}


\subsection{The \textsc{Clumpy} torus properties vs. SED coverage}\label{sec:covSED}

\begin{figure}[ht!]
\includegraphics[angle=0,scale=0.47]{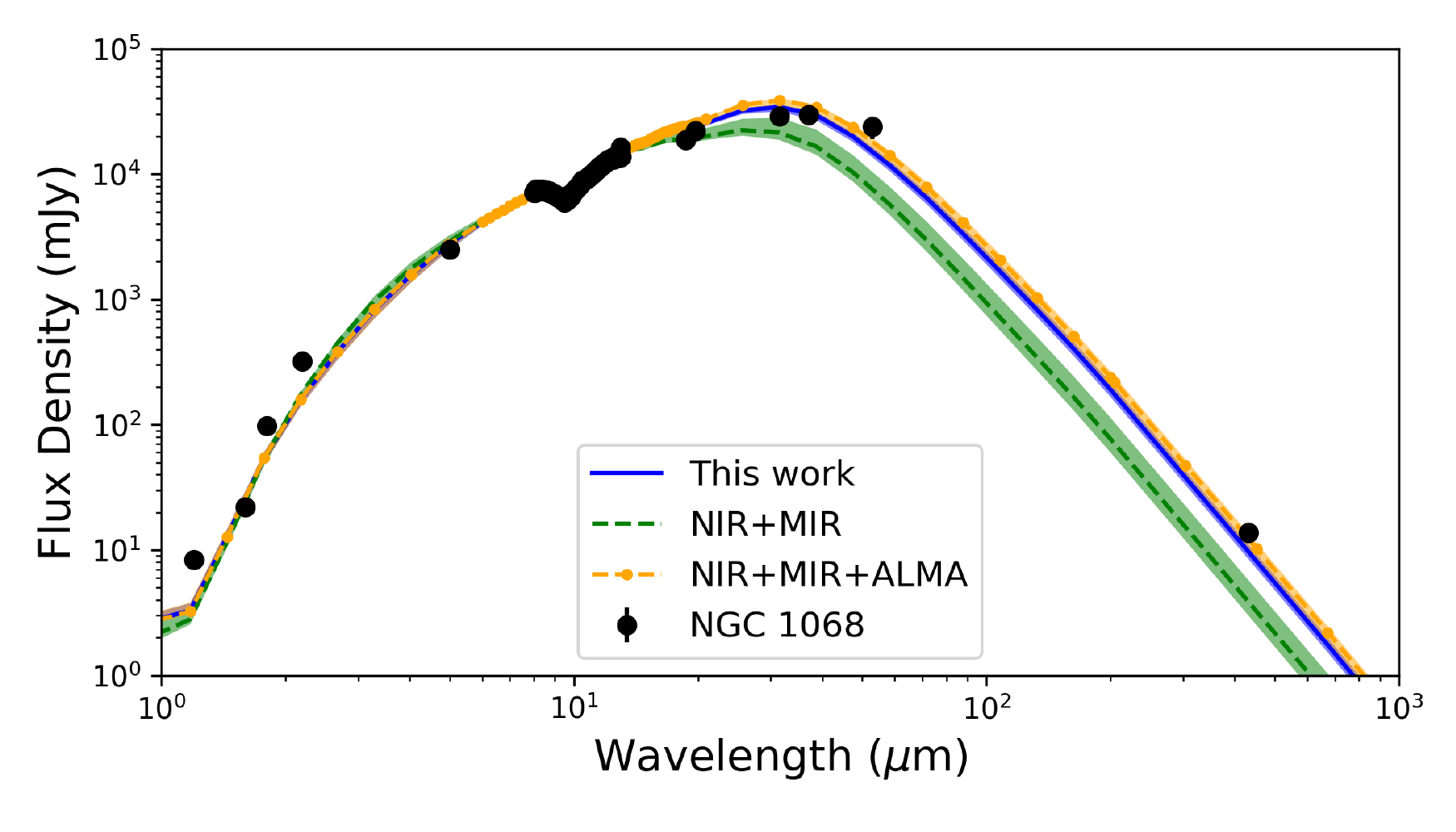}
\caption{\textsc{Clumpy} torus models inferred using different SED sampling (Table \ref{tab:table3}). The posterior distributions for each model are shown in Table \ref{tab:table3} and Figure \ref{fig:fig7}. 
\label{fig:fig5}}
\epsscale{1.}
\end{figure}

We here investigate the emission and distribution of dust in the torus of NGC 1068 using several SED coverages. Direct comparison between previous studies are not straightforward due to the development of the \textsc{clumpy} torus models through the past several years\footnote{News update of \textsc{Clumpy} torus models: \url{https://www.clumpy.org/pages/news-updates.html}}. Our nuclear SED is constructed using the mentioned previous studies in Section \ref{sec:Ctor}. Thus, to avoid any potential misinterpretation of the physics of the torus that slightly different version of the models can introduce, we here re-analyze the nuclear SED as a function of the SED coverage using the most updated version of the \textsc{Clumpy} torus models. Table \ref{tab:table3} shows the output parameters of the best inferred \textsc{Clumpy} torus model for a SED using $1-20$ \um~imaging and spectroscopic observations (labeled as NIR+MIR), and for a SED using $1-20$ \um~imaging and spectroscopic observations and ALMA observations (labeled as NIR+MIR+ALMA). The posterior distributions for each parameter with their median value and 1-$\sigma$ error are shown in Section \ref{app:pos}. For each SED coverage, Fig. \ref{fig:fig5} shows the best inferred \textsc{clumpy} torus model and their 1-$\sigma$ uncertainty.

When \textsc{clumpy} torus model fitting is used with data only in the 1$-$20 \um~wavelength range, 1) the torus is smaller and more compact (large $q$ values) than the current resolved torus of $\sim$10 pc diameter of NGC 1068 by the ALMA observations, and 2) the turn-over of the torus emission peaks at shorter wavelengths than when the SED coverage includes observations at longer wavelengths. This result implies that 1$-$20 \um~observations are not  able to probe the full extent of the torus. Despite the angular resolution, $2.4\arcsec-4.9\arcsec$, Fig. \ref{fig:fig5} shows that the turn-over of the torus emission occurs in the range of 30-40 \um, which corresponds to a characteristic temperature of 70$-$100K. This result indicates that 1) the amount of cold dust, and/or 2) the radiation from indirectly radiated clouds is substantial to shift the peak emission of the torus towards longer wavelengths.


\subsection{2D \textsc{clumpy} torus images}\label{sec:2Dtor}


\begin{figure*}[ht!]
\includegraphics[angle=0,scale=0.43]{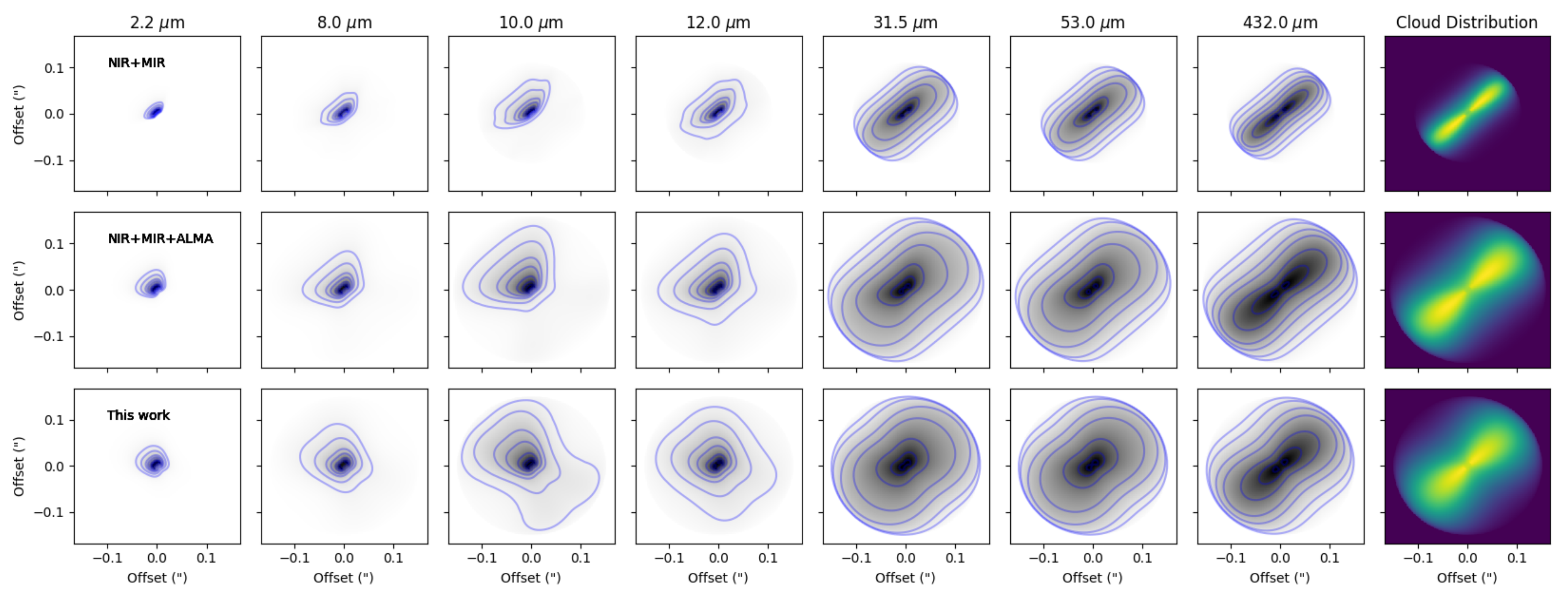}
\caption{2D \textsc{Clumpy} torus images of NGC 1068 generated using HyperCAT based on the several SEDs in Table \ref{tab:table3} as a function of wavelength. The first seven columns show the dust emission from 2.2 \um~to 432~\um, while the last column shows the cloud distribution. Contours shows the intensities at the levels of 0.05, 0.1, 0.2, 0.3, 0.5, 0.7 and 0.9 times the peak flux. In all cases, model was scaled to a distance of 14.4 Mpc, and North is up and East is left.
\label{fig:fig6}}
\epsscale{1.}
\end{figure*}


We use the radiative transfer code \textsc{clumpy} torus \citep{Nenkova:2008ab} to compute the surface brightness and cloud distributions of the dusty torus as a function of wavelength for each set of parameters shown in Table \ref{tab:table3}. Specifically, we use the HyperCubes of AGN Tori (\textsc{HyperCAT}\footnote{\textsc{clumpy} images can be found at \url{https://www.clumpy.org/pages/images.html}}, Nikutta et al. in preparation). \textsc{HyperCAT} uses the \textsc{clumpy} torus models with any combination of parameters to generate physically scaled and flux calibrated 2D images of the dust emission and distribution for a given AGN. We use a distance of 14.4 Mpc, a torus orientation on the plane of the sky of $\sim$138$^{\circ}$ East of North based on the IR polarimetric signature of the nucleus \citep[e.g.][]{Packham:1997aa,Simpson:2002aa,Lopez-Rodriguez:2015aa,Gratadour:2015aa}, and  the bolometric luminosities and model parameters by each inferred model from Table \ref{tab:table3}. Figure \ref{fig:fig6} shows the dust emission distribution from 2.2 \um~to 432 \um~for the $1-20$ \um~SED+MIR Spectroscopy (labeled as NIR+MIR), $1-20$ \um~SED+MIR Spectroscopy+ALMA (labeled as NIR+MIR+ALMA), and this work,  whose torus parameters are summarized in Table \ref{tab:table2}. The last column shows the cloud distribution for each \textsc{clumpy} torus.

For all cases, the clouds are distributed in the equatorial plane with major differences in their torus sizes, angular widths, and radial density profile (Fig. \ref{fig:fig5}-last column). These differences affect the morphology of the dust emission as a function of wavelength. We find that the 2.2 \um~dust emission is concentrated on the inner edge of the torus where dust is directly radiated by the central engine, while the $8-12$ \um~dust emission is along the polar direction as the high opacity in the equatorial direction is absorbing most of the radiation from the central engine. At longer wavelengths, $>$30 \um, the dust emission is along the equatorial plane, where the 432 \um~truly describes the bulk of dust distribution in the torus.

We point out the tight morphological and size similarities between our 2D \textsc{clumpy} torus image at 432 \um, using the well sampled SED from 1 \um~to 432 \um, with the observed torus emission by the ALMA observations \citep{Garcia-Burillo:2016aa,2018arXiv180106564I}. However, when ALMA observations are compared with the 2D images produced by the $1-20$ \um~SED, the inferred torus model is smaller, more compact and thinner, which supports the discussion above regarding the $1-20$ \um~observations underestimates the true size of the torus.

Using our 2D \textsc{clumpy} torus image at 12.0 \um, we find that the dust emission along the polar direction is $\sim$5 times that of the central dust emission. We estimate the dust emission along the polar direction using an ellipse of semimajor axis of 139 mas (9.7 pc) and an eccentricity of 0.91, while the central dust emission was estimated in a circular aperture of 50 mas (3.5 pc) diameter. At this wavelength, the north emission is more prominent than that coming from the south, due to the inferred torus inclination of 75$^{\circ}$. We estimate that the fractional contribution along the polar direction increases at larger inclinations, although further extinction by the host galaxy and/or dust in the narrow line region will play an important role in the observed emission as a function of the location in the galaxy. \cite{Lopez-Gonzaga:2014aa}, using IR interferometric observations with MIDI/VLTI, found that dust emission in the polar regions at scales of $5-10$ pc contributes four times more at 12 \um~that dust located in the torus. They also found a dependency in flux density as a function of the location along the north-south direction attributed to extension by the host galaxy. Our results using a solely 2D \textsc{clumpy} torus model and those by IR interferometric observations are of comparable order of magnitude, which tentatively indicates that the IR interferometric observations of NGC 1068 may be observing the dust emission from the optically thin dust of the torus.


\section{Spectral Decomposition} \label{sec:seddes}

Diffuse extended emission from dust and/or star formation regions surrounding the AGN can  contribute at some level within the unresolved core of the SOFIA observations. What is the contribution of AGN emission within the unresolved core of our SOFIA observations?

We perform an SED analysis and fit the SED of the nuclear emission using the aperture and PSF-scaling photometric measurements, hereafter referred as ``large'' and ``small'' aperture SEDs, respectively.


\begin{figure*}[ht!]
\includegraphics[angle=0,scale=0.40]{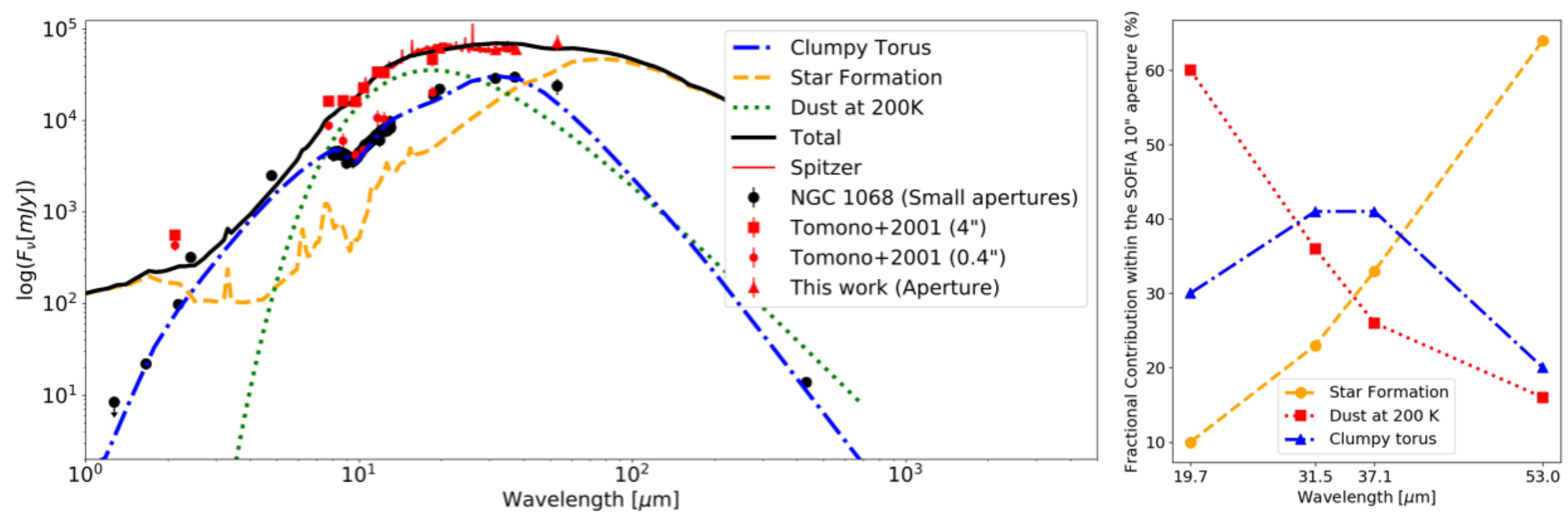}
\caption{Left: Spectral decomposition of the nuclear SED of NGC 1068. The large apertures (red dots and lines) were fitted using star formation region (orange dashed line), \textsc{clumpy} torus (blue dot-dashed line) and a black body component at 200 K (green dotted line). The small aperture photometry and spectroscopy (black dots) was fitted as described in Section \ref{sec:Ctor}. The total model (black line) is shown. Right: Fractional contribution to the total flux within the SOFIA 10\arcsec~aperture from the star formation (orange dashed line/circle), dust at 200 K (red dotted  line/square) and \textsc{clumpy} torus (blue dot-dashed line/triangle) components.
\label{fig:fig4}}
\epsscale{1.}
\end{figure*}


The large aperture SED is composed of our 10\arcsec~aperture photometry (F$_{T}$ in Table \ref{tab:table1}) in combination with \textit{Spitzer} spectroscopic data taken from CASSIS\footnote{The Combined Atlas of Sources with Spitzer IRS Spectra (CASSIS) is a product of the IRS instrument team, supported by NASA and JPL: \url{http://cassis.sirtf.com}} \citep{Lebouteiller:2011aa}. In addition, we also include the $2-20$ \um~photometry in a 4\arcsec~(280 pc) aperture using the Mid-Infrared Test Observation System (MIRTOS) on the 8.2 m Subaru Telescope by \citet{Tomono:2001aa}. Figure \ref{fig:fig4} shows the nuclear SED using large (red dots) and small (black dots) apertures. It is worth noticing that 1) all the photometric measurements using large apertures are consistent with our aperture photometry, F$_{T}$, and 2) our 10\arcsec~photometric measurement at 53 \um~shows an increase in flux density with respect to the $30-40$ \um~photometric measurements, which indicates an extra emissive component at long wavelengths. 

The large aperture SED was fitted as the contribution of the best inferred of the \textsc{clumpy} torus model to the small aperture SED (Section \ref{sec:Ctor}) and a star formation component. We use the \textsc{clumpy} torus model as it better reproduces the smaller aperture SED than the smooth torus models. We use the empirical template of M82 as the star formation component from the \textit{Spitzer}-space-telescope, Wide-field, InfraRed Extragalactic (SWIRE) template library\footnote{SWIRE templates can be found at: \url{http://www.iasf-milano.inaf.it/~polletta/templates/swire_templates.html}} \citep{Polletta:2007aa}. We estimate the minimum reduced $\chi^{2}$ ensuring that the total model was within 10\% of the measured total flux density of the large aperture SED.  We find that the aperture photometric measurements in the $20-53$ \um~wavelength range can be explained by the contribution of the torus emission and star formation region.  However, we have to include an extra component to explain the excess of emission in the $8-20$ \um~wavelength range. This excess emission can be explained with the combination of the torus emission and an additional blackbody component with a characteristic temperature at 200 K. We interpret the dust component at 200 K as dust emission arising from the narrow line region (NLR) in the central 10\arcsec~(700 pc) of NGC 1068  \citep[see][]{Tomono:2001aa}.  Table \ref{tab:table1} lists the fractional contribution of each component within the 10\arcsec~aperture for the SOFIA observations. We estimate a 5\% uncertainty for the fractional contribution of each component shown in Table  \ref{tab:table1}.

Based on our spectral decomposition within the 10\arcsec~(700 pc) nuclear aperture, the fractional contribution to the total flux of the star formation increases with increasing wavelength, from $10\pm1$\% at 19.7 \um~to $64\pm3$\% at 53 \um~(Fig. \ref{fig:fig4}-right). The dust emission from extended dusty structures modeled as a blackbody component with a characteristic temperature at 200 K decreases with increasing wavelength, from $60\pm3$\% at 19.7 \um~to $16\pm1$\% at 53 \um. This extended emission, not associated with the torus, contributes $>$80\% of the total flux in the $8-20$ \um~wavelength range and it is attributed to the N-S dust emission as seen by \cite{2000AJ....120.2904B}, also previously suggested by \citet{1993ApJ...419..136C,Mason:2006aa}. The fractional contribution to the total flux of the torus emission shows a turn-over in the range of $30-40$ \um~with a maximum fractional contribution to the total emission of $41\pm2$\%, reaching a minimum of $20\pm1$\% at 53 \um.

We can compare the potential contribution from the torus emission within the PSF-scaling photometry estimated in Section \ref{subsec:pho}. In general, the fractional contribution of the total flux from the PSF-scaling method,  \% PSF in Table \ref{tab:table1}, is slightly larger than the torus emission estimated by the spectral decomposition, \%Torus in Table \ref{tab:table1}. Specifically, the PSF-scaling method agrees with the torus emission within the PSF of SOFIA at all wavelengths within a fraction of $\sim$10\% in the $20-53$ \um~wavelength range. Based on Fig. \ref{fig:fig4}, the turn-over of the torus emission in the range of 30$-$40 \um~can be distinguished from a) the expected peak emission at $\sim$100 \um~by star formation regions, and b) extended dust emission associated with the NLR at shorter wavelengths.


\section{Conclusions} \label{sec:con}

Using SOFIA observations taken with FORCAST ($19.7-37.1$ \um) and HAWC+ at 53.0 \um~onboard SOFIA, we observationally find the turn-over of the torus emission of NGC 1068 to be in the $30-40$ \um~wavelength range. Specifically, the torus emission increases from 1 \um~to 30 \um~and then we measure a decrease in the unresolved nuclear emission at 53 \um~with respect to the photometric measurements in the $30-40$ \um~wavelength range. This result is in agreement with the observational constraint that the turn-over does not occur until wavelengths $>$31.5 \um~found by \cite{Fuller:2016aa} using a sample of 11 Seyfert galaxies. Using \textsc{Clumpy} torus models, we found a radius of $r_{out} = 5.1^{+0.4}_{-0.4}$ pc for the torus of NGC 1068. Our estimation of the torus size is in excellent agreement with the recently observed diameter of $7-10$ pc by ALMA \citep{Garcia-Burillo:2016aa,Gallimore:2016ab,2018arXiv180106564I}. Although smooth torus models produce compatible results with those found by the \textsc{clumpy} torus models, they overestimate the nuclear SED in the FIR wavelength range. Despite the angular resolution of SOFIA, the $20-53$ \um~SOFIA observations together with 1) PSF-scaling and spectral decomposition techniques, and 2) \textsc{clumpy} torus models provide a tool to characterize the size of the torus in a large sample of AGN when sub-arcsecond resolution observations by ALMA is not available.

We computed 2D images for the best inferred \textsc{clumpy} torus model using several SED coverages. We found that the full extent and the cold dust of the torus are underestimated when a nuclear SED covering the $1-20$ \um~wavelength range is used. The inferred \textsc{clumpy} torus from our $1-432$ \um~nuclear SED reproduces well the ALMA observations. Specifically, the dust emission at 432 \um~is spatially coincident with the cloud distribution of the torus, while the morphology of the dust emission in the $1-20$ \um~wavelength range probes mostly optically thin dust located above and below the equatorial plane of the torus. We estimated a contribution of the polar dust emission at 12 \um~to be $\sum$5 times that from the central source, which indicates that the IR interferometric observations of NGC 1068 may be observing the dust emission from the optically thin dust of the torus.

\acknowledgments

Based on observations made with the NASA/DLR Stratospheric Observatory for Infrared Astronomy (SOFIA). SOFIA is jointly operated by the Universities Space Research Association, Inc. (USRA), under NASA contract NAS2-97001, and the Deutsches SOFIA Institut (DSI) under DLR contract 50 OK 0901 to the University of Stuttgart. Financial support for this work was provided by NASA through award \#02\_0035  and \#04\_0048 issued by USRA. E.L.-R. acknowledges support from the Japanese Society for the Promotion of Science (JSPS) through award PE17783, the National Observatory of Japan (NAOJ) at Mitaka and the Thirty Meter Telescope (TMT) Office at NAOJ-Mitaka for providing a space to work and great collaborations during the short stay at Japan. A.A.-H. acknowledges financial support from the Spanish Ministry of Economy and Competitiveness through grant AYA2015-64346-C2-1-P which is party funded by the FEDER program. C.R.A. acknowledges the Ram\'on y Cajal Program of the Spanish Ministry of Economy and Competitiveness through project RYC-2014-15779 and the Spanish Plan Nacional de Astronom\'ia y Astrofis\'ica under grant AYA2016-76682-C3-2-P.

%

\vspace{5mm}
\facilities{SOFIA(FORCAST, HAWC+)}


\software{astropy \citep{2013A&A...558A..33A},  
          }



\appendix
\section{\textsc{clumpy} torus model paramaters posteriors}\label{app:pos}

\begin{figure*}[ht!]
\includegraphics[angle=0,scale=0.7]{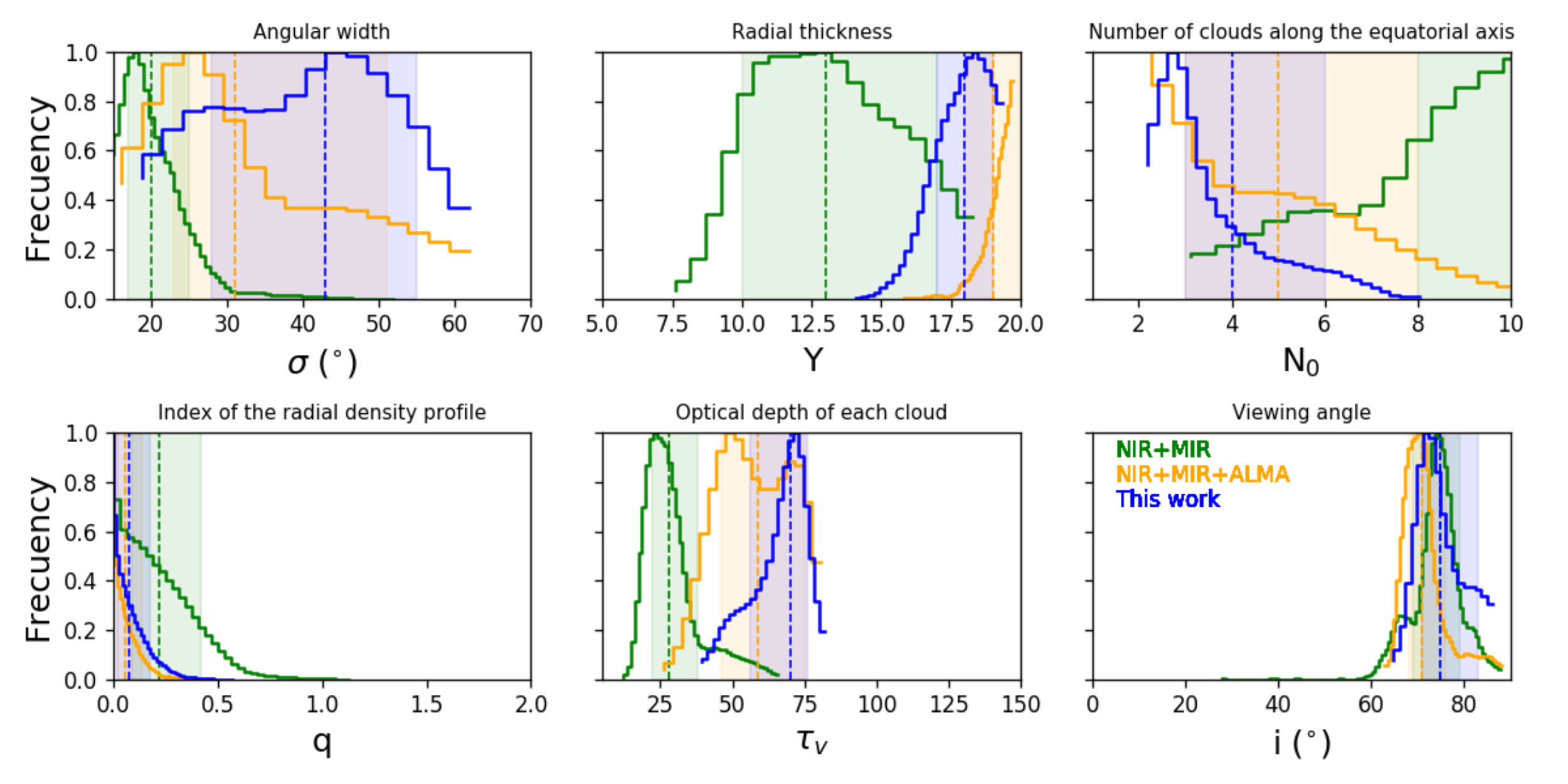}
\caption{Posterior distributions (black lines) of the \textsc{clumpy} torus model parameters for the best model shown in Fig. \ref{fig:fig3} and \ref{fig:fig5}. The blue shadowed region shows the 1$-\sigma$ uncertainty and the blue line shows the best inferred model. 
\label{fig:fig7}}
\epsscale{1.}
\end{figure*}

Figure \ref{fig:fig7} shows the posterior distributions of the best inferred \textsc{clumpy} torus model shown in Fig. \ref{fig:fig3} and \ref{fig:fig5}.


\bibliography{ngc1068_sofia}






\end{document}